%% file: Clustering_JSA.tex
\documentclass[12pt, letterpaper]{article}

\AtBeginDocument{
  \providecommand\BibTeX{{
    \normalfont B\kern-0.5em{\scshape i\kern-0.25em b}\kern-0.8em\TeX}}}

\usepackage{geometry}
\newgeometry{vmargin={25mm}, hmargin={25mm,25mm}}

\usepackage{physics}
\usepackage{xcolor}
\usepackage{pifont} 
\usepackage{colortbl}
\usepackage{tabularx}
\usepackage{amsmath}
\usepackage{mathtools}
\usepackage{caption}
\usepackage{subcaption}
\usepackage{booktabs}

\usepackage{multirow}
\usepackage{listings}
\usepackage[ruled]{algorithm2e}

\usepackage{authblk}
\usepackage{url}

\usepackage{cite}

\definecolor{mGreen}{rgb}{0,0.6,0}
\definecolor{mGray}{rgb}{0.5,0.5,0.5}
\definecolor{mPurple}{rgb}{0.58,0,0.82}
\definecolor{mOrange}{rgb}{0.9,0.4,0.1}
\definecolor{backgroundColour}{rgb}{0.95,0.95,0.92}

\lstdefinestyle{CStyle}{
    backgroundcolor=\color{backgroundColour},   
    commentstyle=\color{mGreen},
    keywordstyle=\color{magenta},
    numberstyle=\tiny\color{mGray},
    stringstyle=\color{mPurple},
    basicstyle=\scriptsize,
    breakatwhitespace=false,         
    breaklines=true,                 
    captionpos=b,                    
    keepspaces=true,                 
    numbers=left,                    
    numbersep=5pt,                  
    showspaces=false,                
    showstringspaces=false,
    showtabs=false,                  
    tabsize=2,
    frame=single,
    rulecolor=\color{black},
    language=C
}

\newcommand{\upcid}{$^\ddagger$}
\newcommand{\bscid}{$^\dagger$}

\begin{document}

\title{At-Scale Evaluation of Weight Clustering to Enable Energy-Efficient Object Detection}
\author[\bscid, \upcid]{Martí Caro}
\author[\bscid]{Hamid Tabani}
\author[\bscid]{Jaume Abella}
\affil[\bscid]{Barcelona Supercomputing Center (BSC)}
\affil[\upcid]{Universitat Polit\`{e}cnica de Catalunya (UPC)}

\date{}

\maketitle

\begin{abstract}

Accelerators implementing Deep Neural Networks (DNNs) for image-based object detection operate on large volumes of data due to fetching images and neural network parameters, especially if they need to process video streams, hence with high power dissipation and bandwidth requirements to fetch all those data.
While some solutions exist to mitigate power and bandwidth demands for data fetching, they 
are often assessed in the context of limited evaluations with a scale much smaller than that of the target application, which challenges finding the best tradeoff in practice.

This paper sets up the infrastructure to assess at-scale a key power and bandwidth optimization -- weight clustering -- for You Only Look Once v3 (YOLOv3), a neural network-based object detection system, using videos of real driving conditions. Our assessment shows that accelerators such as systolic arrays with an Output Stationary architecture turn out to be a highly effective solution combined with weight clustering. In particular, applying weight clustering independently per neural network layer, and using between 32 (5-bit) and 256 (8-bit) weights allows achieving an accuracy close to that of the original YOLOv3 weights (32-bit weights). Such bit-count reduction of the weights allows shaving bandwidth requirements down to 30\%-40\% of the original requirements, and reduces energy consumption down to 45\%.
This is based on the fact that (i) energy due to multiply-and-accumulate operations is much smaller than DRAM data fetching, and (ii) designing accelerators appropriately may make that most of the data fetched corresponds to neural network weights, where clustering can be applied.
Overall, our at-scale assessment provides key results to architect camera-based object detection accelerators by putting together a real-life application (YOLOv3), and real driving videos, in a unified setup so that trends observed are reliable.
\end{abstract}

\input{1.0.Introduction}

\input{2.0.Background}
\input{3.0.Modelling}
\input{4.0.Evaluation}
\input{5.0.Related}
\input{6.0.Conclusions}

\section*{Acknowledgements}
The DRAC project under grant 001-P-001723 is co-financed by the European Union Regional Development Fund within the framework of the ERDF Operational Program of Catalonia 2014-2020 with a grant of 50\% of total eligible cost. This work has also been partially supported by the Spanish Ministry of Science and Innovation under grant PID2019-107255GB-C21/AEI/10.13039/501100011033.

\bibliographystyle{unsrt}
\bibliography{cites.bib}

\end{document}

%% file: 1.0.Introduction.tex
\section{Introduction}
\label{sec:intro}

Computing platforms deliver increasing levels of performance over time, which has allowed deploying performance-hungry functionalities across many applications and domains. For instance, camera-based object detection is used in a plethora of applications building on deep neural networks (DNNs), whose accuracy already matches the needs of many of those applications~\cite{Goodfellow-et-al-2016,DBLP:journals/corr/Schmidhuber14}, with realizations such as ResNet-101~\cite{ResNet}, ResNet-152~\cite{ResNet}, YOLO~\cite{yolo} and YOLOv3~\cite{yolo-v3}.
Detection accuracy and computing cost (e.g., in the form of number of arithmetic operations needed) have also been used as key parameters to compare DNNs~\cite{yolo-v3}. However, other parameters such as memory bandwidth and energy consumption are also key to adopt specific DNN implementations.

Due to efficiency, DNN applications are often implemented using specialized accelerators that devote most resources (e.g. area and power) to do effective computing work such as performing multiply-and-accumulate (MAC) operations, which have been shown to be the basis of DNNs~\cite{DLmatmul}. Those accelerators include systolic arrays~\cite{1163741}, GPUs~\cite{10.1145/1281500.1281643}, as well as other application-specific designs~\cite{MOOLCHANDANI2021101887}. However, in all cases, those accelerators require fetching large amounts of data that include (i) the image to be processed, and (ii) the weights of the DNN implemented~\cite{intro-cnns}. Generally, this implies that large volumes of data need to be fetched from DRAM memory where data is stored.

DRAM energy consumption has been shown to be the main source of energy consumption whenever data locality is poor and DRAM accesses are frequent~\cite{10.1145/2541940.2541967}, hence making the computing unit (e.g. the accelerator) require large and sustained amounts of data being fetched from DRAM memory. In the particular case of camera-based object detection,
DNNs with some acceptable degree of accuracy processing large enough images (e.g. as in the case of autonomous driving) may easily need to fetch some gigabytes of data to handle each image, being most of it due to the weights that need to be fetched for the DNN~\cite{yolo-v3}. Hence, data locality cannot realistically be achieved since images to be processed change every few milliseconds, and the weights that need to be reused across images require too much space (i.e. hundreds of MBs) to store them in any local cache or register file. Therefore, reducing the amount of data to be fetched for energy reasons, but also to reduce the memory bandwidth requirements, is a key challenge for the design of systems for camera-based object detection.

Different approaches have been investigated to decrease memory energy and bandwidth requirements of DNNs, being reduced precision and weight clustering two of the most prominent ones. The former, reduced precision~\cite{8114708}, aims at using lower precision arithmetic, and hence, lower precision data representations, to decrease the amount of data needed. For instance, using 32-bit floating point numbers instead of 64-bit ones halves the amount of data needed. On the other hand, reducing precision further discretizes the weights of the DNN, hence, impacting accuracy. Therefore, it can be used to some extent.
The latter, weight clustering~\cite{8114708}, is orthogonal to the former and 
aims at restricting the number of different weights to a limited number (e.g., 128 different values) while keeping their precision (e.g., each of those 128 values is a 32-bit floating point number) so that those high-precision weights need being fetched just once (e.g., 128 32-bit values), and the weights used by the DNN are encoded with fewer bits (e.g., 7 bits per weight instead of 32) to select the particular full-precision weight that must be used in each case. While reduced precision could be used instead of weight clustering matching the number of precision bits (e.g., 7-bit weights) to the bits used for clustered weights, both of them would have almost identical memory bandwidth requirements (e.g., 7 bits per weight both, plus 128 32-bit values for weight clustering), but weight clustering provides much higher precision (e.g., 32-bit weights instead of 7-bit ones) and hence, better prediction accuracy~\cite{8114708}.

The energy savings achieved with weight clustering have only been illustrated with limited examples, but, to our knowledge, the bandwidth reduction and energy savings of weight clustering have not been evaluated in realistic (at scale) case studies.

This paper addresses this challenge by assessing the gains that can be achieved with weight clustering in terms of memory bandwidth and energy in the context of a real-size application, the YOLOv3 camera-based object detector~\cite{yolo-v3}, which we assess with real driving videos for its use in the context of autonomous driving.
In particular, the contributions of this work are as follows:
\begin{itemize}
\item We integrate different power models for the MAC units, the DRAM memory, and the SRAMs needed by weight clustering in a consistent framework to estimate the energy consumption of the DRAM and accelerator devices with and without weight clustering.
\item We integrate YOLOv3 with the power models to estimate the energy consumption to process each frame from a video.
\item We assess the tradeoff between overall energy, memory bandwidth and prediction accuracy of weight clustering using both, reference labelled data sets and videos from real driving conditions. Our evaluation considers both, global and per-layer clustered weights, and different numbers of weight clusters, from 32 (5 bits) up to 256 (8 bits).
\end{itemize}
Our results show that weight clustering can be effectively used to preserve detection accuracy while decreasing significantly memory bandwidth requirements and overall energy in the context of autonomous driving. In particular, per-layer weight clusters show to be particularly effective to preserve accuracy unaffected with 7-bit and 8-bit clusters decreasing DRAM energy consumption by around 60\% w.r.t. the original 32-bit data used by YOLOv3. Our results show that, with negligible accuracy loss, 5-bit clusters allow reaching DRAM energy reductions of around 68\%. Bandwidth also drops from 200 GB/s for 25 frames-per-second (FPS) down to 60-80 GB/s depending on the bits used for clustering.

The rest of the paper is organized as follows. Section~\ref{sec:back} provides some background on DNNs, appropriate accelerators and weight clustering. Section~\ref{sec:model} presents the case study. Section~\ref{sec:eval} evaluates the case study. Related work is provided in Section~\ref{sec:related}. Section~\ref{sec:concl} summarizes this work.

%% file: 2.0.Background.tex
\section{Background}
\label{sec:back}

This section provides some background on DNNs in general, and Convolutional Neural Networks (CNNs) in particular. It also provides some details on dataflow accelerators for DNNs, and introduces clustering and its application to DNN weights.

\subsection{DNNs}

\begin{figure}[!h]
	\centering
	\includegraphics[width=0.75\columnwidth]{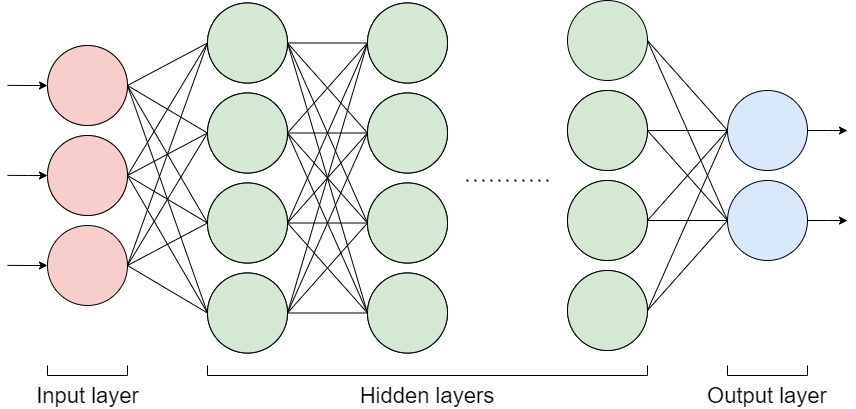}
	\caption{Structure of a DNN.}
	\label{fig:dnn-structure}
\end{figure}

DNNs may include tens or even more than one hundred hidden layers, which is not generally the case for other types of neural networks.
Figure~\ref{fig:dnn-structure} illustrates the general structure of a DNN. DNNs consist of an input layer with $n$ inputs, an output layer with $m$ outputs and $k$ hidden layers in-between, where $n$, $m$ and $k$ are at least $1$. Each node is called a \textit{neuron} and its value is computed with the values of the neurons in the previous layer that are connected to it, the weights associated to these connections and a bias. The weights and biases are learned in the training phase. In the model shown in Figure~\ref{fig:dnn-structure}, all neurons are fully connected across layers, however, this is not always the case.

Formally, a neuron is computed as presented in Equation \ref{eq:neuron}, where $n_{i,j}$ denotes a neuron in position $j$ of layer $i$, $w_{i,k,j}$ is the weight of the connection between $n_{i-1,k}$ and $n_{i,j}$, $m_{i-1}$ denotes the number of neurons in layer $i-1$, and $b_{i,j}$ is the bias of $n_{i,j}$.

\begin{equation} \label{eq:neuron}
	n_{i,j} = b_{i,j} + \sum_{k=1}^{m_{i-1}} n_{i-1,k}\times w_{i,k,j} 
\end{equation}

CNNs are one of the most popular types of DNNs. CNNs architecture encodes image-specific features allowing it to be primarily used for finding patterns in images to recognize objects. CNNs are comprised of an input layer that holds the pixel values of the image, convolutional layers that perform convolutions of the inputs with the weights and biases, pooling layers that downsample the spatial dimensionality of the input, fully-connected layers that produce class scores, and an output layer that holds the final classification of the objects~\cite{intro-cnns}.

\begin{figure}[!h]
	\centering
	\includegraphics[width=0.75\columnwidth]{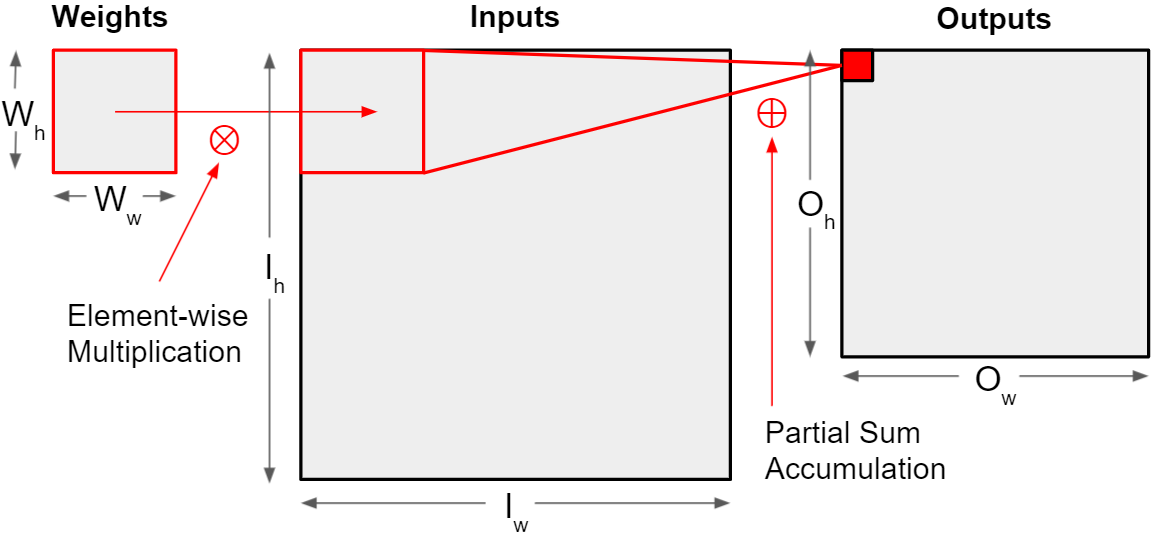}
	\caption{2-D convolution process.}
	\label{fig:convolution}
\end{figure}

Figure~\ref{fig:convolution} shows the traditional 2-D convolution used for image processing. The weights consist of an $W_h \times W_w$ matrix, the inputs of an $I_h \times I_w$ matrix, and the outputs of an $O_h \times O_w$ matrix. The output height $O_h$ can be calculated as $O_h = \frac{W_h - I_h + 2P}{S} + 1$, where $W_h$ is the number of rows of the weights, $I_h$ is the number of rows of the inputs, $P$ is the padding value (i.e. the number of rows and columns added to the frame of the image if needed, to cover the entire image with the sliding window), and $S$ is the stride value (i.e. the number of positions to move the sliding window). The output width is calculated using the number of columns instead of the number of rows, $O_w =  \frac{W_w - I_w + 2P}{S} + 1$.

The process of a 2-D convolution is performed with a sliding window in which the weight window moves through the entire input, performing an element-wise multiplication and accumulation. In the example shown, we can see that the weights are multiplied by a $W_h \times W_w$ subset of the input and accumulated to obtain a single output element. The window then moves through the entire image to compute all the $O_h \times O_w$ outputs.

\subsection{YOLOv3 Object Detection System}
YOLOv3~\cite{yolo,yolo-v3} is a camera-based detection software framework capable of processing images efficiently in real-time. YOLOv3 implements one of the DNNs delivering higher accuracy, and hence, it is very popular and has already been adopted as part of several industrial applications, including Apollo~\cite{apolloAuto}, which is an industrial autonomous driving framework.
YOLOv3~\cite{yolo-v3} is capable of detecting around 80 different classes of objects, including multiple animals, common objects and food, but also multiple types of vehicles, pedestrians and road signs, with the latter three groups being of particular relevance for autonomous driving. 

For each image processed, YOLOv3 detects the objects in it and attaches each one a confidence level describing the confidence on the detection and classification of such object (i.e., a value between 0\% and 100\%). However, a confidence threshold is used to drop those objects whose confidence is below the threshold (e.g., 50\%), hence keeping only those for which the detection has been performed with sufficiently high confidence. 

We use Darknet framework~\cite{darknet-framework}, developed by YOLO authors, implementing YOLOv3 model for image and video processing.
For the sake of reliability, Darknet averages confidence levels across a specific number of images in the case of video processing (e.g., across 3 images). This allows mitigating false positives and false negatives that occur in just one frame, since their sporadic and erroneous confidence is averaged with that of the two previous frames. For that purpose, the average for each value of the YOLO layer output of the current image and the YOLO layer output of the previous two images is calculated.

Finally, YOLOv3 provides two different models; Tiny YOLOv3, which is a lighter and simplified version of YOLOv3 that can run over 200 frames per second, but at the cost of lower accuracy, and the baseline YOLOv3, which is a more sophisticated model with longer execution time but also more accurate. In this work we have used the baseline YOLOv3, which provides the highest accuracy and is, generally, more relevant for industry.

\subsection{Accelerators for DNNs}
\label{sec:acc}

\begin{figure}[!h]
	\centering
	\includegraphics[width=.75\columnwidth]{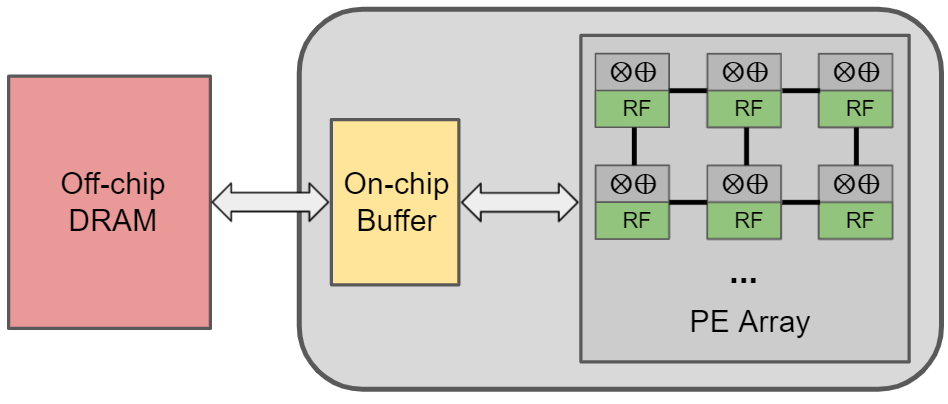}
	\caption{Overview of a CNN accelerator architecture.}
	\label{fig:accelerator}
\end{figure}

In this work, our reference accelerator is assumed to have an architecture such that its energy per MAC is low (i.e., most power dissipation occurs in MAC units), and whose utilization can be kept high (e.g., close to 100\%) in the case of DNNs. These two characteristics are the basis of a high-performance and power-efficient accelerator for DNNs.

Systolic arrays~\cite{SystolicArrays} adhere to those characteristics. Their typicall architecture is depicted in Figure~\ref{fig:accelerator} for illustration purposes. As shown, such an accelerator includes an array of processing elements (PEs) capable of performing MACs (e.g., a MAC per cycle if properly pipelined). In particular, those PEs multiply inputs and weights, and accumulate the result with the output.
Each PE is normally implemented with its own local (and small) register file (RF) or scratchpad memory, which is used to retain data to be reused so that data locality can be effectively exploited minimizing costly data transactions. Generally, those RFs allow retaining one of the three operands of the MAC, whereas the other two need being fetched for each operation from a slower and higher power memory (DRAM in our case), or forwarded by other PEs.

Depending on what data is kept in the PE-local RFs, different dataflows can be defined. For instance, a \emph{weight stationary} dataflow keeps the weights to be operated in those RFs with the aim of not having to fetch them, hence maximizing their reuse and reducing the energy consumption devoted to fetch weight. An \emph{output stationary} dataflow, instead, stores in the RF the results of the MACs to be accumulated with the results of the following MACs to be executed. Such architecture intends to minimize the energy consumption needed to read and write output results. Other possible dataflows -- although less popular -- include \emph{input stationary} dataflows to minimize the energy devoted to fetch inputs, \emph{row stationary} dataflows that aim at reducing the energy devoted to fetch all data types simultaneously (i.e., inputs, weights and outputs), and \emph{no local reuse} dataflows that do not reuse any data, but allow removing RFs from the PEs.

As shown later in this work, we consider an Output Stationary dataflow, as it is a popular dataflow that can benefit significantly from using weight clustering.

\subsection{Weight Clustering}

Weights of a DNN typically are the largest set of data to fetch, and so it is in the case of YOLOv3, which requires roughly 8 GB of weights to be fetched from memory to process each single image. This poses, obviously, high pressure on the memory bandwidth to allow real-time operation. For instance, if a value of 25 FPS is wanted -- typical for video processing -- around 200 GB/s of memory bandwidth are required. This is an overwhelming cost for autonomous driving systems, both in terms of the hardware procurement costs to allow such a high bandwidth, and in terms of power to read so much data per second. Note that autonomous driving systems employ multiple cameras and lots of other sensors which increase the bandwidth and computation requirements manifold resulting in significant pressure on the memory subsystem.

This work assesses the effectiveness of a quantization technique called Weight Clustering or Sharing to mitigate the cost of fetching weights for DNNs. In particular, we assess its impact in both, accuracy and energy consumption. Weight clustering has been shown to be an effective solution to decrease the size of the data to be operated with very limited impact on accuracy. Weight clustering builds on limiting the number of different weights to be used, which are represented with fewer bits while preserving the original precision of the weights. This is achieved by making multiple weights being identical by using the mean value of a cluster (i.e. a group of weights), with those weights being stored in a centroid table (also known as codebook table).
K-means~\cite{k-means,k-means-2} is the most popular clustering algorithm. This algorithm works by finding the similarity between the data and grouping the data in $K$ groups called clusters. 

Using this technique, the FP32 values of the weights used by YOLOv3 that are stored in a DRAM are replaced by smaller integer indices that are used to access the centroid table. Note that this table may be orders of magnitude smaller than the weights and can fit in a small on-chip SRAM within the systolic array. This adds the extra cost of performing an indirection, but since the indirection fetches data from a tiny on-chip SRAM with a significantly shorter access time than the DRAM, the expected overhead is negligible and clustering is expected to significantly reduce the power and memory bandwidth used to fetch the weights from memory. In terms of delay, weight translation through the centroid table can be organized to occur ahead of time so that full 32-bit weights are available in the PEs whenever needed.
For instance, if we set $K=256$ clusters for K-means, we will require only 256 different FP32 weights for the DNN, and we will be able to encode the weights of the DNN with only 8 bits to identify the particular FP32 weight to use for each one of them. The centroid table will have to store 256 unique FP32 values, which is only 1KB of data. Note that the size of the centroid table is independent of the number of weights of the DNN.

\begin{figure}[!h]
	\centering
	\includegraphics[width=1\columnwidth]{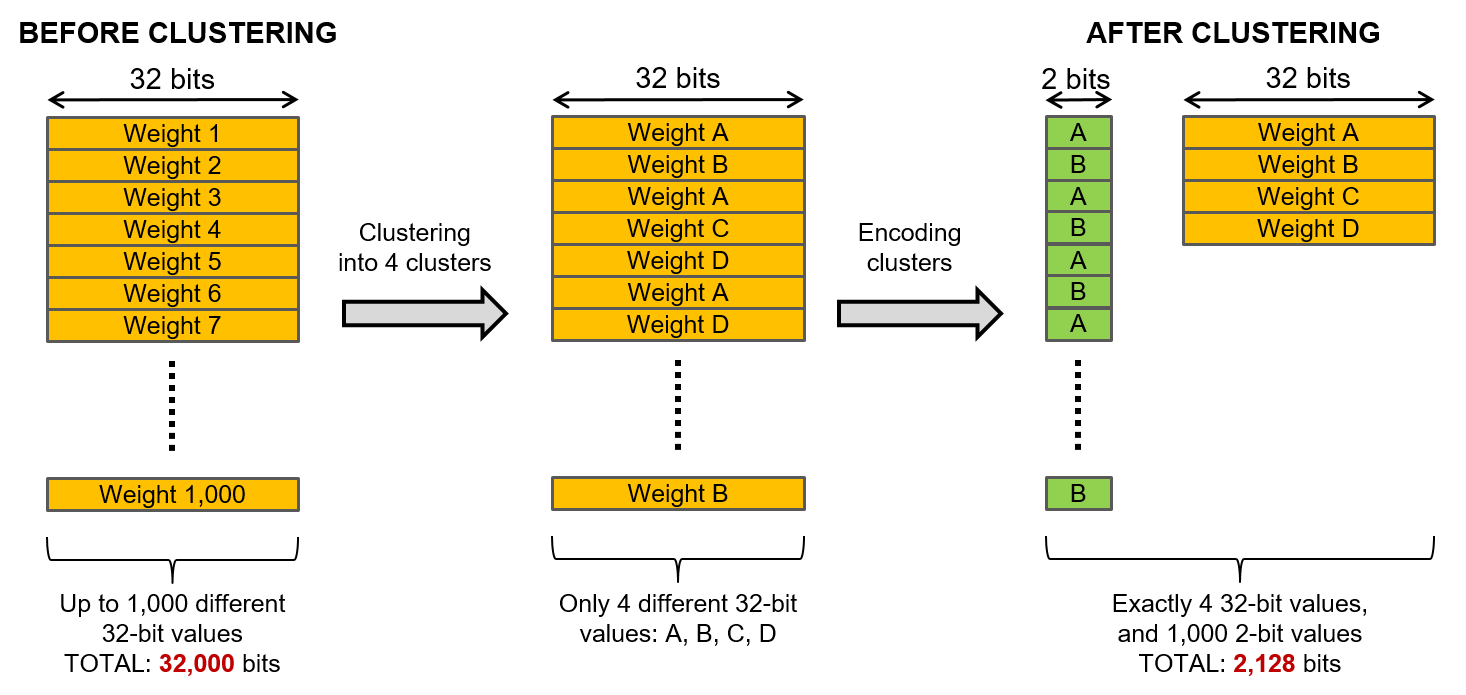}
	\caption{Example where 1,000 32-bit weights are clustered into 1,000 2-bit weights.}
	\label{fig:exampleclus}
\end{figure}

For the sake of illustration, we provide a small example where we have 1,000 32-bit (e.g., FP32) weights, see left side of Figure~\ref{fig:exampleclus}. Those require 32,000 bits of storage. However, using K-means, in this example, we identify 4 32-bit values that can be used for clustering the 1,000 potentially different 32-bit values. We refer to those 4 clustered weights as \emph{weight A}, \emph{weight B}, \emph{weight C}, and \emph{weight D} in the figure. As shown in the central part of the figure, we replace each one of the original 32-bit weights by the representative weight of the cluster where the corresponding weight has been mapped to with K-means. This intermediate step is only shown for illustrative purposes, but not applied in practice. Finally, we replace each of the 1,000 32-bit weights by a 2-bit index indicating to what of the 4 32-bit values such weight has been mapped. Therefore, as shown in the rightmost part of the figure, we need 1,000 2-bit values to store the clustered weights and the translation table with the 4 32-bit weights, 2,128 bits in total, which corresponds to a 15x storage reduction in this example.

Note that using reduced precision, as mentioned before, leads to similar memory bandwidth requirements (2,000 bits in the example). However, using lower precision constrains the accuracy of the DNN. In the example, we could only use 2-bit values for the weights. This reduces the cost of the arithmetic, but restricting precision impacts accuracy if the number of bits used becomes too low. Instead, weight clustering keeps similar bandwidth requirements to those of comparable precision reduction (e.g., 2,128 bits vs 2,000 bits in the example), but allows using weights with much higher precision (e.g., 32-bit weights), which are operated with the appropriate arithmetic units (e.g., FP32 arithmetic in the example) and hence, allows the DNN reaching much higher accuracy than just using lower precision~\cite{han2016deep,fp-energy,DBLP:journals/corr/ChoiEL16,9122495,ye2018unified,gong2014compressing,app9122559}.

%% file: 3.0.Modelling.tex
\section{Case Study Modelling}
\label{sec:model}

This section presents the design and modelling choices for our case study. Those choices relate to the underlying design of the systolic array, the actual application of weight clustering, the power model, and the input data used for evaluation.

\subsection{Systolic Array}
As discussed in Section~\ref{sec:acc}, systolic arrays can be organized in different ways depending on what data is fetched, and what data remains stationary -- if any. 
Given that weight clustering allows a drastic reduction of the bandwidth and energy requirements related to fetching weights, it is expected that the most appropriate dataflow is the one where any other type of data (inputs and outputs) requires low bandwidth and energy. Otherwise, Amdahl's law would severely limit the achievable gains. Therefore, a systolic array with an output stationary dataflow is the most appropriate baseline organization to consider. Such an organization allows keeping intermediate results local in the PEs until they are final, hence not needing to fetch them. Weights, instead, must be fetched every time they are required, and they account for a much larger size than input data operated, potential gains with weight clustering are huge if we use an output stationary dataflow since the number of bits per weight that must be fetched from memory is drastically reduced.

Moreover, despite not being explicitly evaluated in our study, output stationary systolic arrays minimize write operations to memory since intermediate results are not sent out. If the memory space where those results are stored is coherent, write operations may trigger coherence messages into the cores, potentially harming performance. Hence, minimizing those write operations may also bring benefits in terms of reduction of coherence messages triggered.

\subsection{Weight Clustering Application}

In our case study, we consider two alternative ways to realize weight clustering:
\begin{itemize}
\item \emph{All layers} (global). In this case, clustering is applied using as input all weights of all layers at once. This approach minimizes the number of weights needed, but may impact accuracy since all layers are enforced to use the same set of weights.
\item \emph{Per layer} (local). Clustering is applied separately to the weights of each layer individually. Accuracy is gained since each layer may use the best set of weights for the corresponding layer. On the other hand, weights need to be loaded every time a new layer starts. Note, however, that the cost of fetching few floats (e.g. between 32 and 256 floats) per layer is negligible compared to the typical tens of MBs of data needed per layer.
\end{itemize}

To implement both alternatives, we have used the K-means algorithm, as indicated before, and modified the relevant functions of YOLOv3 to read and process the weights accordingly. Listing~\ref{code-clustering} shows the changes made to the convolution operation, showing the added indirection needed to fetch the weights. Line 10 corresponds to the original weight read operation, and line 11 to the operation using clustering. Note that the example corresponds to the simple case with 8-bit clustered weights. We have used also lower numbers of bits (5, 6 and 7), whose code is similar to the case of 8 bits, but require some bit manipulation given that they are not byte aligned.

\begin{table}[!h]
\begin{lstlisting}[style=CStyle, caption={C implementation of gemm function using clustered weights.}, label={code-clustering}]
void gemm_nn_centroids(int M, int N, int K, float ALPHA, 
  float *A, int lda, 
  float *B, int ldb, 
  float *C, int ldc, 
  float *centroids, unsigned char *indexes)
{
  int i,j,k;
  for(i = 0; i < M; ++i){
    for(k = 0; k < K; ++k){
      // float A_PART = ALPHA * A[i*lda+k];
      float A_PART = ALPHA * centroids[indexes[i*lda+k]];
      for(j = 0; j < N; ++j){
        C[i*ldc+j] += A_PART * B[k*ldb+j];
} } } }
\end{lstlisting}
\end{table}

K-means has already been considered in the past realize clustering of activations and/or weights~\cite{han2016deep,fp-energy,gong2014compressing,app9122559,DBLP:journals/corr/ChoiEL16,ye2018unified,8578919,9122495,DBLP:journals/corr/CourbariauxB16,DBLP:journals/corr/CourbariauxBD15,wu2020integer,DBLP:journals/corr/ZhouYGXC17,DBLP:journals/corr/abs-1806-08342}. However, to our knowledge, our work is the first one assessing the benefits, in terms of energy and bandwidht, of post-training weight clustering for an at-scale case study such as YOLOv3.

\subsection{Power Model}

For our assessment, we have modelled the main power contributors of an accelerator running the DNN, namely DRAM accesses and MAC operations~\cite{8114708}. Since centroids tables are direct overheads of weight clustering, we have also included their energy consumption modelling them as SRAM tables. Instead, other components of the accelerator, such as its peripheral logic, have not been included as their contribution to the total energy of the accelerator is relatively small and, in any case, it will not experience significant variations regardless of whether weight clustering is used or not.

\subsubsection{Modelled DRAM and SRAM specifications}

In our at-scale assessment of YOLOv3, we must take into account two type of memories: DRAM memory where data is stored and fetched from, and the centroids tables (SRAM memories) used when weight clustering is implemented. Both types of memories have been modelled using the CACTI tool~\cite{cacti}, which models energy consumption, as well as area and delay, for a wide range of memories.
Hence, it allows collecting DRAM and SRAM energy estimates building on comparable models.
Tables~\ref{tb:dram-spec} and ~\ref{tb:sram-spec} describe the specific parameters used to model DRAM and SRAM memories respectively in our case study. Regarding the technology node considered, it has been chosen to keep consistency with the energy values available for MAC operations, taken from~\cite{fp-energy}, as described later. This allows modelling properly the contribution to total energy of each component, and hence, obtaining results where the relative cost of DRAM and MAC energy consumption is kept consistent and trends are meaningful, despite absolute values do not correspond to the latest technology nodes in the market.

\begin{table}[!h]
      \caption{Modelled DRAM configuration.}
      \centering
      \label{tb:dram-spec}
	\begin{tabular}{|l||c|c|}
		\hline
		\multicolumn{2}{|c|}{\textbf{DRAM Configuration}} \\
		\hline
		\textbf{Type:} & DDR4-3200 \\
		\hline
		\textbf{Size (GB)} & 1 \\
		\hline
		\textbf{Channels} & 8 \\
		\hline
		\textbf{Bus width (bits)} & 64 \\
		\hline
		\textbf{Burst length (bits)} & 64 \\
		\hline
		\textbf{Technology (nm)} & 45 \\
		\hline
		\textbf{Page size (KB)} & 1 \\
		\hline
		\textbf{Energy per read (pJ)} &  2937\\
		\hline
		\textbf{Energy per write (pJ)} &  2937\\
		\hline
		\textbf{Interface frequency (GHz)} &  3.2\\
		\hline
	\end{tabular}
\end{table}

\begin{table}[!h]
      \centering
      \caption{Modelled SRAM configuration.}
      \label{tb:sram-spec}
	\begin{tabular}{|l||c|c|}
		\hline
		\multicolumn{2}{|c|}{\textbf{SRAM Configuration}} \\
		\hline
		\textbf{Bus width (bits)} & 32 \\
		\hline
		\textbf{Burst length (bits)} & 32 \\
		\hline
		\textbf{Technology (nm)} & 45 \\
		\hline
	\end{tabular}
\end{table}

For DRAM accesses, the energy reported in Table~\ref{tb:dram-spec} corresponds to 8-byte random accesses. However, an output stationary dataflow exploits spatial reuse in the memory row buffers~\cite{dnn-reuse}. Hence, assuming 1 KB row buffers, we can expect 1 row miss every 128 accesses, while the remaining 127 accesses are row hits, whose energy consumption is significantly lower. Since CACTI does not provide data for row hits, we have used DRAMsim3~\cite{dramsim3} to estimate the relation between row hits energy consumption and row misses one. For that purpose, we have collected energy measurements for a sequence of one of the default DRAM configurations (DDR4\_1Gb\_x16\_1866) and we have used two traces to estimate the energy ratio of row hits and misses. To do so we have executed read and write access traces with spatial locality (1 miss followed by 127 hits) and without spatial locality (all misses), and extrapolated the energy consumption of hits w.r.t. misses. Results of each access type are shown in Table~\ref{tb:dram-avg-energy}.

\begin{table}[!h]
	\caption{DRAM Average Read and Write Energy.}
	\label{tb:dram-avg-energy}
	\setlength{\tabcolsep}{1.0mm} 
	\centering
	\begin{tabular}{|l||c|c|c|}
		\hline
		Row Miss Read Energy (pJ) &  2937\\
		\hline
		Row Hit Read Energy (pJ) &  1735\\
		\hline
		\textbf{Average Read Energy (pJ)} &  1753\\
		\hline
		\hline
		Row Miss Write Energy (pJ) &  2953\\
		\hline
		Row Hit Write Energy (pJ) &  1859\\
		\hline
		\textbf{Average Write Energy (pJ)} &  1876\\
		\hline
	\end{tabular}
\end{table}

Regarding the SRAM memory for centroids, its size and energy vary depending on the clustering configuration as reported in Table~\ref{tb:sram-energy}. Note that, in this case, the reported SRAM read and write energy is for fetching 32 bits of data, since weights for YOLOv3 are 32-bit floating point numbers. The read ports number of the SRAM memory has been set large enough to allow translating as many weights as fetched per cycle in each case. In any case, SRAM energy cost is negligible regardless of its number of ports and size since it is several orders of magnitude smaller than that of DRAM accesses.
Regarding SRAM static energy, it has also been included but it is completely negligible even if we assume a physical SRAM per layer as potentially needed in the per-layer clustering. In fact, very few tables would be needed if we load centroids right before they are needed, but as said, SRAM static energy is simply irrelevant (below 0.1\% of the total energy in all cases).

\begin{table}[!h]
	\caption{SRAM read energy for each cluster configuration.}
	\label{tb:sram-energy}
	\setlength{\tabcolsep}{1.0mm} 
	\centering
	\begin{tabular}{|l||c|c|c|}
		\hline
		\textbf{Configuration} & \textbf{SRAM size (bytes)} & \textbf{Read Energy (pJ)} \\
		\hline
		\textbf{Cluster-8 bits} & 1024 & 0.85\\
		\hline
		\textbf{Cluster-7 bits} & 512 & 0.52\\
		\hline
		\textbf{Cluster-6 bits} & 256 & 0.40\\
		\hline
		\textbf{Cluster-5 bits} & 128 & 0.36\\
		\hline
	\end{tabular}
\end{table}

\subsubsection{Number of Memory Accesses}

Memory access counts have been computing building on the feature maps of the weights, outputs, and inputs, their padding and stride, and the dataflow of the accelerator (Output Stationary in our case), as well as considering spatial data reuse. We consider the memory accesses of all layers in the neural network used by YOLOv3 for object detection, namely, the convolutional, route, upsample, shortcut, and YOLO layers.

We note that YOLOv3 has 3 types of weight filters: 3x3 with Stride 1, 3x3 with Stride 2, and 1x1 with Stride 1. Next, we specify the formulas for calculating the number of memory accesses based on the three types of weights, where appropriate. The meanings of the abbreviations used in the formulas are, W: Weights, I: Inputs, O: Outputs and the meaning of the subscripts, w: width, h: height, c: channels, and f: filters. 

For the number of memory accesses, it is assumed that each access fetches/writes a single FP element (32 bits). Knowing that the modelled DRAM has a bus width of 64 bits and that the read/write energy reported is for 64-bit accesses, we divide the number of accesses by 2 for DRAM accesses.

\begin{itemize}
	\item Convolutional Layer: Performs the convolutions between weights and inputs.
    	\begin{itemize}
        		\item Weight Reads (3x3, Stride 1, Stride 2) = $W_w \times W_h \times W_c \times W_f \times (I_h-2)$
        		\item Weight Reads (1x1, Stride 1) = $W_w \times W_h \times W_c \times W_f \times I_h$
	 	\item Input Reads (3x3, Stride 1) =  $I_w \times W_h \times W_c \times (I_h-2)$
		\item Input Reads (3x3, Stride 2) = $(I_w+1) \times W_h \times W_c \times (I_h-2)$
		\item Input Reads (1x1, Stride 1) = $I_w \times W_h \times W_c \times I_h$
		\item Output Writes = $O_h \times O_w \times W_f$
   	 \end{itemize}
	\item Shortcut Layer: Add the feature maps of two layers.
    	\begin{itemize}
        		\item Reads: $I_h \times I_w \times I_c + I_{h'} \times I_{w'} \times I_{c'}$
		\item Writes: $I_h \times I_w \times I_c + I_{h'} \times I_{w'} \times I_{c'}$
   	 \end{itemize}
	\item Route Layer: Outputs a specific layer or concatenates two layers.
    	\begin{itemize}
        		\item Reads One Layer: $I_h \times I_w \times I_c$
		\item Reads Concatenation: $I_h \times I_w \times I_c + I_{h'} \times I_{w'} \times I_{c'}$
        		\item Writes One Layer: $I_h \times I_w \times I_c$
		\item Writes Concatenation: $I_h \times I_w \times I_c + I_{h'} \times I_{w'} \times I_{c'}$
   	 \end{itemize}
	\item Upscale Layer: Upscales the output of a layer (2x upscaling in our case).
    	\begin{itemize}
        		\item Reads: $I_h \times I_w \times I_c$
		\item Writes: $4 \times I_h \times I_w \times I_c$
   	 \end{itemize}
	\item YOLO Layer: Performs the object detections.
    	\begin{itemize}
        		\item Reads: $I_h \times I_w \times I_c$
		\item Writes: $I_h \times I_w \times I_c$
   	 \end{itemize}
\end{itemize}

\subsubsection{Number of FP operations and their energy estimation}

The number of FP operations has been computed using different procedures for convolutional layers and the other layers. For convolutional ones, we use the following formula to compute the number of MACs:
$$
\#MACs = O_w \times O_h \times W_w \times W_h \times W_c \times W_f
$$
To calculate the number of FP operations from other layers or sources (such as the activations), we have used software counters. The FP operations used during object detection are the following: addition (FPADD), multiplication (FPMUL), subtraction (FPSUB), division (FPDIV), exponential (FPEXP), and square root (FPSQRT). 

Energy consumption of FP operations has been obtained from~\cite{fp-energy} in the case of FPADD and FPMUL, for a 45nm technology node, which, as discussed before, is in synch with the one sued for our memory models (DRAM and SRAM) to perform fair energy comparisons. However, we lack explicit energy estimates for FPDIV, FPSUB, FPSQRT and FPEXP. Nevertheless, those operations represent a negligible part of the FP operations performed by YOLOv3 since the MACs of the convolutional layer already account for 99.54\% of the overall FP operations in YOLOv3. Hence, we have modelled those other operations, which represent a fraction of the remaining 0.46\% FP operations, considering that their energy consumption matches that of an FPMUL operation.

\subsection{Datasets}

For our case study, we have used two different types of datasets: a labelled one, and a set of real driving videos.

A labelled dataset is composed by a set of images labelled by humans. This means that, for each image, the bounding boxes of the objects have been drawn and classified as accurately as possible by a human. These accurate bounding boxes are called ground truths and are used to compare against the predicted bounding boxes of an object detector, to determine its accuracy by using different metrics. Due to its large number of images, and due to defining object categories comparable to those of YOLOv3, we have used the COCO dataset~\cite{coco-detection}. In particular, this dataset contains ten thousand high quality images of complex everyday scenes containing common objects in their natural context. Moreover, the training set of this dataset is the one used to train YOLOv3. The drawback of this dataset is that we are mostly interested in images of road scenes taken from vehicle cameras, but this dataset contains more general contexts. Therefore, we have dropped all those images that do not include any vehicle, therefore keeping 880 images. In any case, note that the fact that those images include vehicles does not imply that they belong to driving scenarios. Hence, the accuracy of the object detection process for this set of images is only relevant in relative terms to compare the accuracy of different setups, but not to obtain conclusions from the absolute values.

We have also considered a set of unlabelled automobile videos to complement the analysis of labelled data. To perform this analysis, we have used thirty seconds extracts, taken at 30 FPS, from six videos of real driving conditions, which provide more relevant input data for your case study~\cite{video-0,video-1,video-2,video-3,video-4,video-6}.
The main disadvantage of using these videos is that we lack a ground truth for comparison purposes since videos are unlabelled. Hence, the only possible comparison is to build the default YOLOv3 without weight clustering as if it was the ground truth. This may lead to some pessimism in terms of accuracy for weight clustering since it is frequent that objects in the boundary between being identified or discarded (e.g. those whose confidence of detection is around 50\%) are relatively often misclassified, being either false positives or false negatives. Hence, a relevant number of misclassified objects in the baseline YOLOv3 may be properly classified with weight clustering -- partly due to the stochastic nature of the object detection process -- but accounted as misclassifications due to the lack of a ground truth. 

In the case of videos, we fully build on the redundancy delivered by YOLOv3 so that, as explained before, object detections are given averaging the confidence level obtained across multiple frames -- 3 in particular -- rather than the confidence values obtained frame by frame. This approach mitigates the risk of sporadic misclassifications.
For instance, an object detected with confidence levels 80\%, 90\% and 40\% across three frames would be regarded as a detection with 70\% confidence, thus above the threshold (e.g. 50\%) despite in the last image it was below the threshold.

\subsection{Practical Integration in an Accelerator}

\begin{figure}[!h]
	\centering
	\includegraphics[width=0.75\columnwidth]{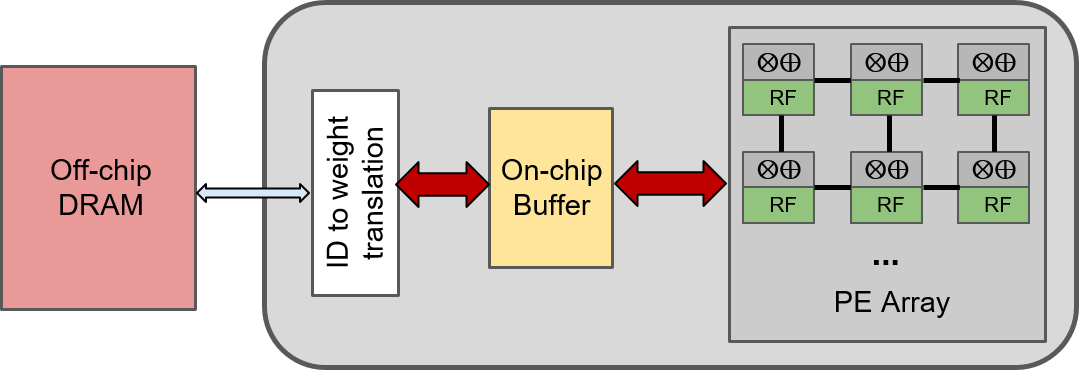}
	\caption{Example of clustering integration at the data fetch interface. Arrows only reflect DNN weights, not image data.}
	\label{fig:integ-out}
\end{figure}

Weight clustering can be integrated in multiple manners in an accelerator. The simplest one consists of a translation layer for weights fetched from memory so that the internal design of the accelerator remains unchanged. This is illustrated in Figure~\ref{fig:integ-out} for a systolic array accelerator, where arrows width indicates bit-width. As shown, the accelerator stores and operates full weights as in a non-weight clustering design. Cluster IDs are fetched from memory and immediately translated into full weights to keep the accelerator design unchanged. This has the advantage of being non-intrusive with the accelerator design. However, this prevents exploiting the fact that cluster IDs require fewer bits than full weights, so that IDs could be propagated further delaying the translation.

\begin{figure}[!h]
	\centering
	\includegraphics[width=0.75\columnwidth]{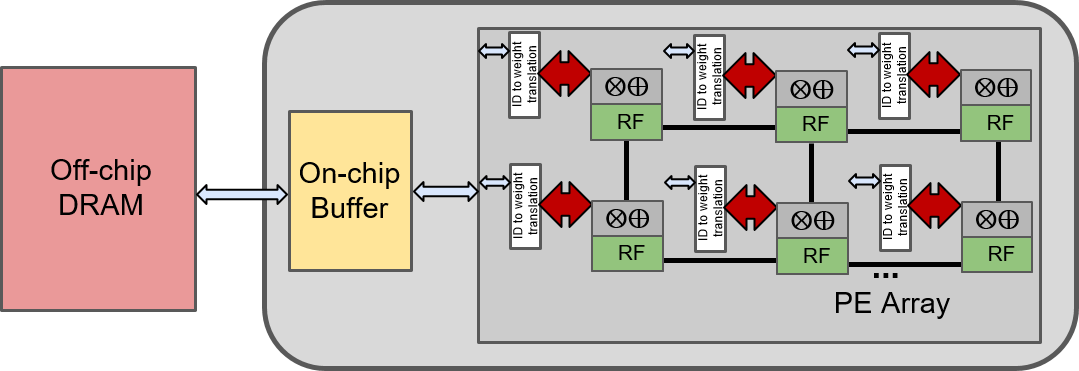}
	\caption{Example of clustering integration at the processing element (PE) interface. Arrows only reflect DNN weights, not image data.}
	\label{fig:integ-in}
\end{figure}

Figure~\ref{fig:integ-in} shows a completely opposed design approach where ID translation occurs as late as possible, right before data are operated in the PEs. This allows column and row management logic (i.e. feeders) use small IDs (e.g. 6-bit) rather than full weights (e.g. 32-bit), as well as narrower connections to transmit weights to PEs. On the other hand, translations may occur in further locations simultaneously, so likely weight translation tables need being replicated across multiple locations. Overall, there is a trade-off between decreasing the width of the weights propagated by using IDs instead of full weights, and the cost of replicating translation logic in multiple locations. 
Note that intermediate solutions where, for instance, translation occurs per column or per row are also possible, therefore providing other trade-offs to choose the most effective one, which is fully accelerator dependent and may change across different accelerator architectures or topologies.

%% file: 4.0.Evaluation.tex
\section{Evaluation}
\label{sec:eval}

This section evaluates the different tradeoffs between accuracy, energy consumption and memory bandwidth for both, the COCO dataset and the real driving videos. 

\subsection{Energy Breakdown}

\begin{figure}[!h]
  \centering
  \begin{minipage}[b]{0.37\textwidth}
    \includegraphics[width=\textwidth]{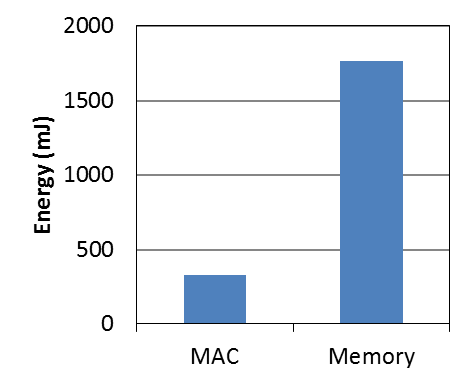}
    \caption{Energy breakdown.}
    \label{fig:energy-usage}
  \end{minipage}
  \hfill
  \begin{minipage}[b]{0.465\textwidth}
    \includegraphics[width=\textwidth]{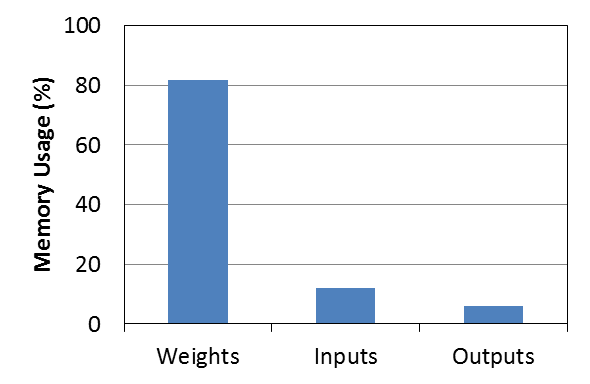}
    \caption{Memory distribution.}
    \label{fig:mem-usage}
  \end{minipage}
\end{figure}

To illustrate the energy contribution of DRAM accesses to fetch DNN weights in the baseline YOLOv3 configuration, we start estimating the energy consumption devoted to both components, namely DRAM accesses and FP operations, with the model described in previous section applied to YOLOv3. In particular, we estimate the energy needed to process one image.
As shown in Figure~\ref{fig:energy-usage}, 2086 mJ are required to process an image, which would correspond to 52 Watts if one image was processed every 40ms (i.e., 25 FPS). As shown, 84.4\% of such energy is devoted to DRAM accesses, which are, therefore, the main energy contributor.

Figure~\ref{fig:mem-usage} shows the distribution of memory accesses of the baseline YOLOv3, dividing the memory accesses into weights (only reads), inputs (only reads), and outputs (reads and writes). 
As shown, 81.9\% of the DRAM accesses correspond to weights, whereas inputs and outputs account for 12.0\% and 6.1\% only respectively. Hence, weights are the main energy contributor of DRAM accesses.

If we put both results together, we come to the conclusion that DRAM accesses to fetch weights almost account for 70\% of the total energy for the considered output stationary systolic array. Hence, weight clustering has significant potential to decrease both, energy consumption and bandwidth requirements.

\subsection{Accuracy Evaluation}

To compare the accuracy of the different configurations considered, we use the Mean Average Precision (mAP)~\cite{padillaCITE2020}. The mAP is used as a standard metric to evaluate and compare the accuracy of object detectors leveraging true positives, false negatives and false positives into a single value (the mAP). Such value allows assessing and comparing the prediction accuracy of object detectors.
The mAP builds upon the Intersection over Union (IoU) metric~\cite{electronics10030279}, which determines whether two bounding boxes overlap enough to regard them as the same object detection. We refer the interested reader to the original works describing those metrics for further details. 
To measure the mAP in our experiments, we use the public implementation provided in~\cite{map-script}.

\begin{figure}[!h]
	\centering
	\includegraphics[width=0.6\columnwidth]{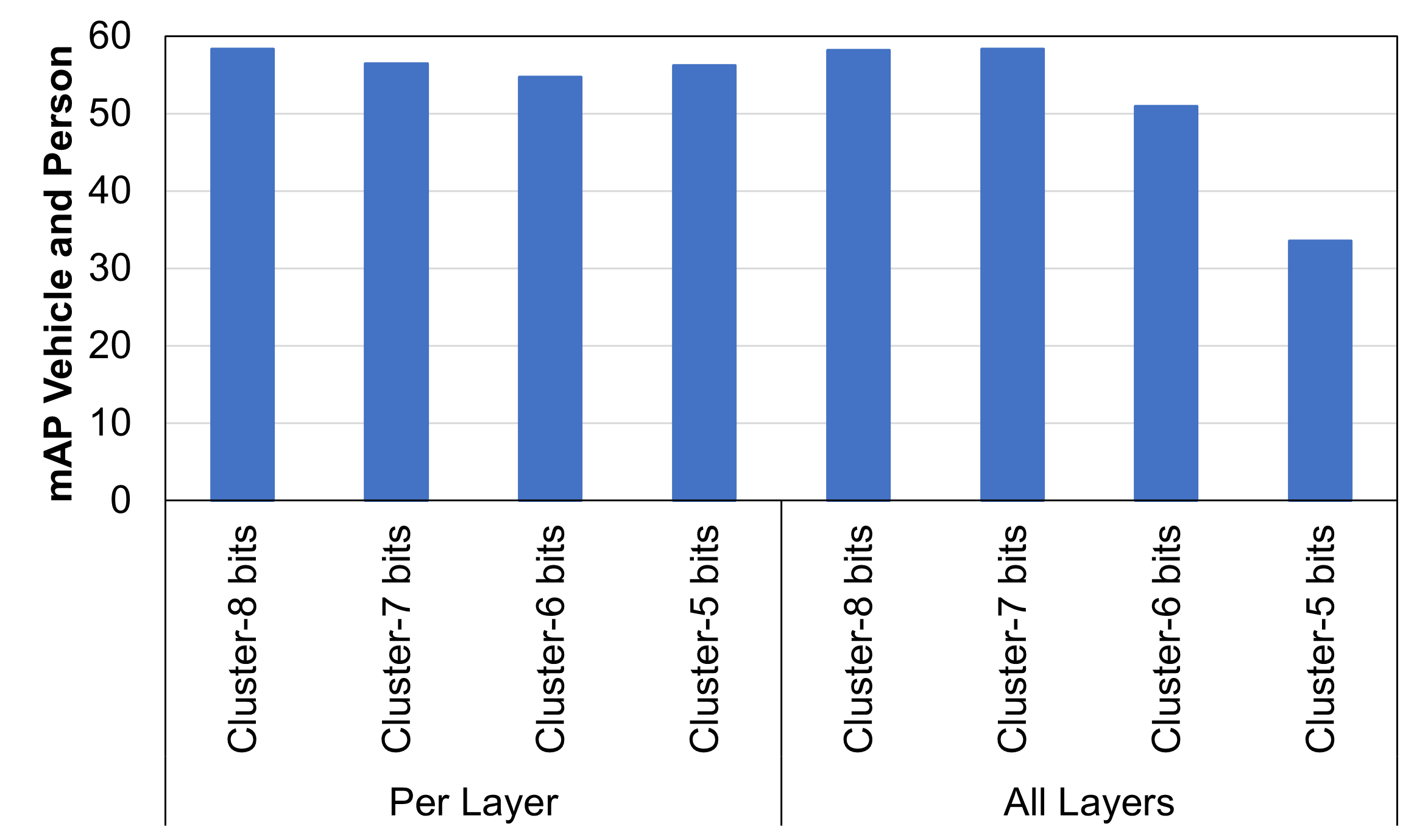}
	\caption{mAP of vehicles and people for the Labelled dataset for the baseline and different clustering configurations.}
	\label{fig:clustering-labelled-vp}
\end{figure}

The results when applying weight clustering for the labelled dataset are shown in Figure~\ref{fig:clustering-labelled-vp}, where we provide the mAP for person and vehicle categories together. Clustering accuracy is given for those configurations providing relevant mAP values, namely between 5 and 8 bits, thus discarding weight clustering with 4 bits or fewer. More than 8 bits have not been considered since 8 bits already achieves mAP values very close to those of the non-clustered YOLOv3.

Results include per-layer clustering (left) as well as global clustering for all layers (right). As shown in the figure, mAP remains high for 7-bit and 8-bit clusters independently of whether clustering is applied for all layers at once or per layer. However, if we further reduce the number of clusters to 64 (6 bits) or 32 (5 bits), then only per-layer clustering achieves acceptable mAP. Instead, with those few clusters, if we apply clustering simultaneously to all layers, accuracy degrades significantly. 

\begin{figure}[!h]
	\centering
	\includegraphics[width=0.6\columnwidth]{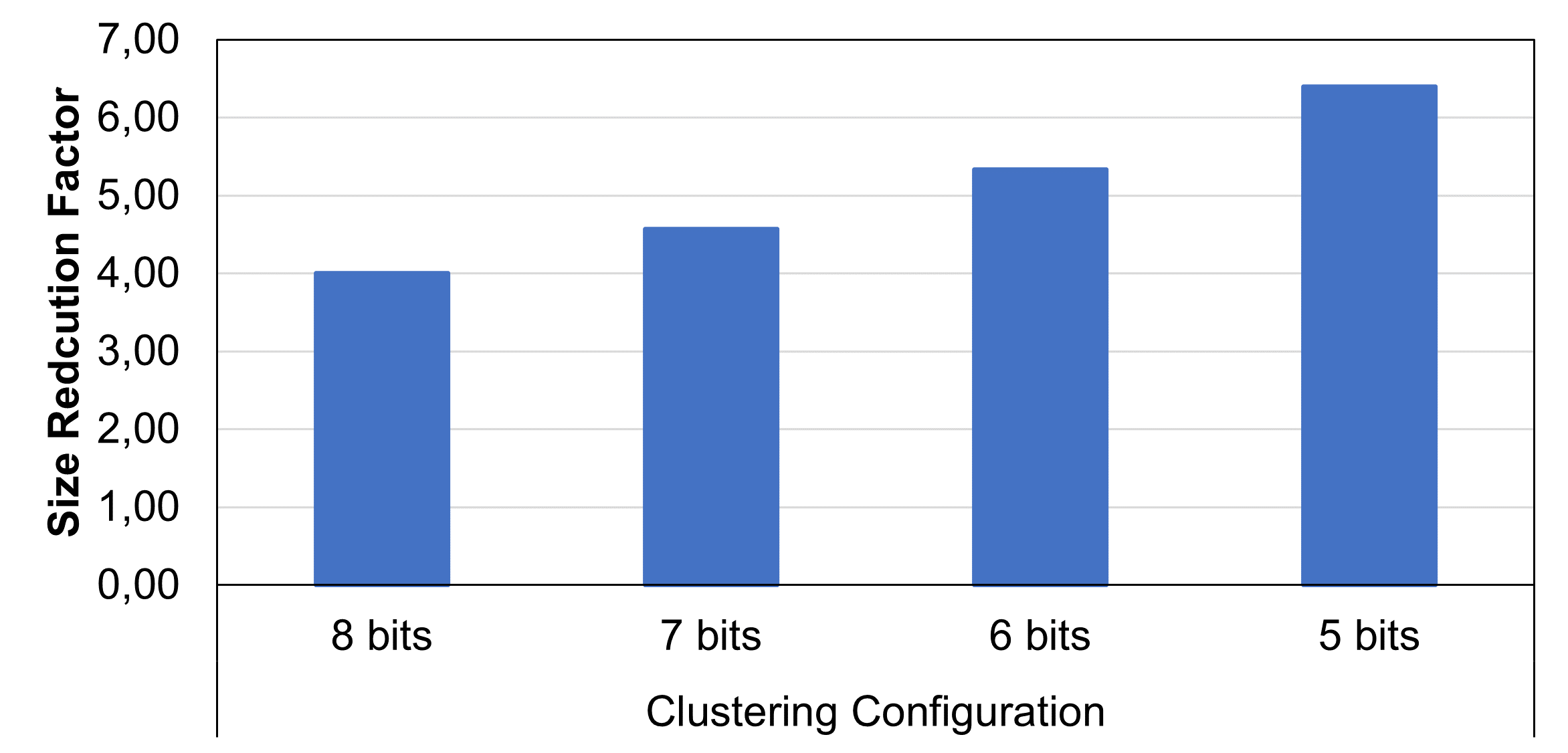}
	\caption{Size reduction factor for the different clustering configurations.}
	\label{fig:size-reduction}
\end{figure}

The reduction of the data size for the different configurations is shown in Figure~\ref{fig:size-reduction}. For instance, 8-bit weight clustering produces a 4x data reduction given that each weight is stored as an 8-bit index rather than a 32-bit FP number in DRAM memory. Note that, if only the labelled dataset is considered, 5-bit clusters per layer offer very good accuracy and a size reduction factor for weights above 6, meaning that original weights (32 bits) require more than 6x memory space than clustered ones (5 bits).

\begin{figure}[!h]
	\centering
	\includegraphics[width=0.6\columnwidth]{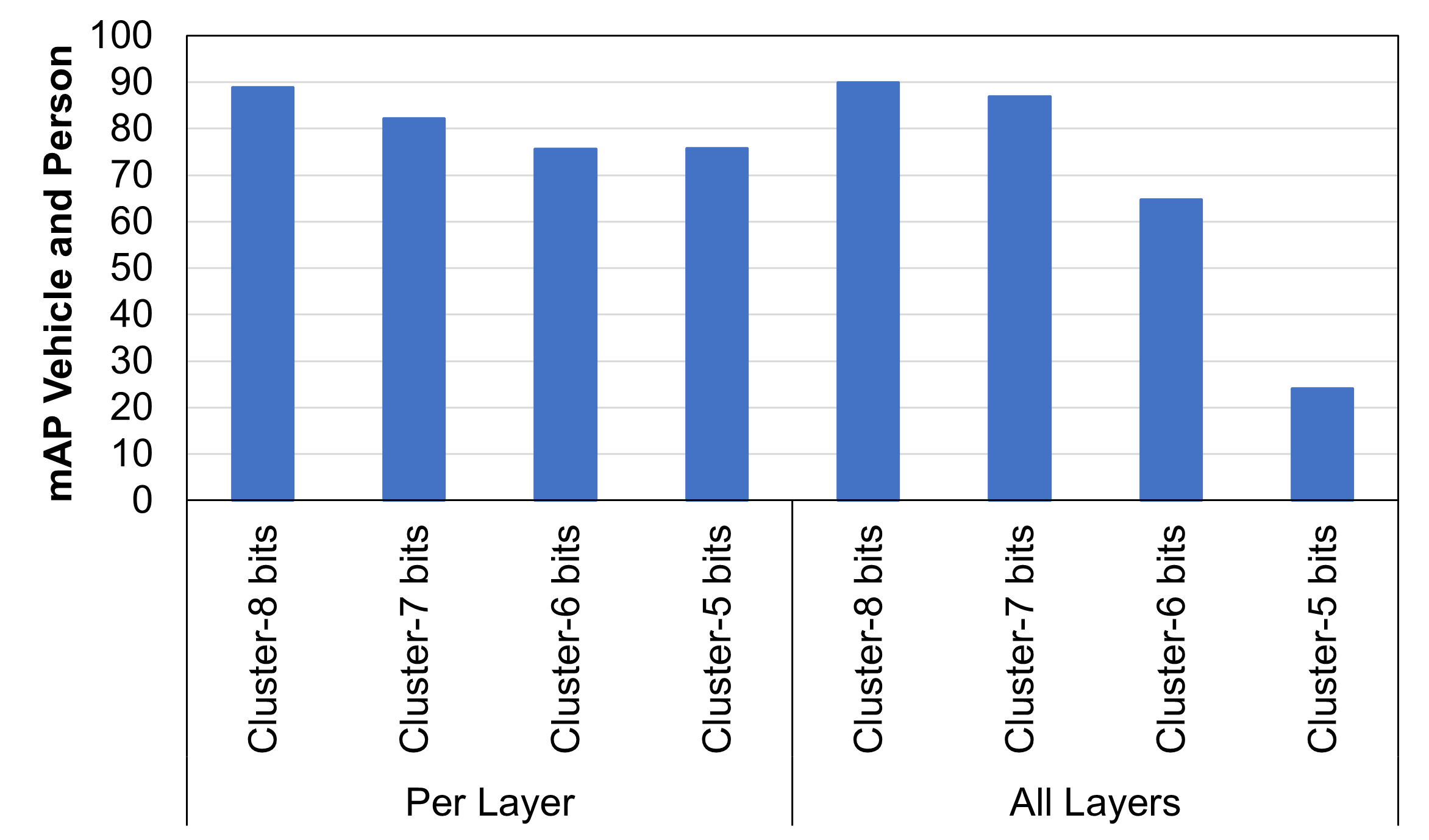}
	\caption{mAP of vehicles and people for the Videos for the baseline and different clustering configurations.}
	\label{fig:clustering-ti}
\end{figure}

Figure~\ref{fig:clustering-ti} shows the mAP for vehicle and person objects together for the videos. First, note that there is some drop in the mAP w.r.t. the baseline. Visual inspection reveals that, as indicated before, this is not real accuracy loss but different object classifications for objects in the boundary. In general, this effect is mostly captured in the mAP drop when using 8-bit clusters. 
Therefore, additional accuracy loss can be confidently attributed to the impact of clustering on accuracy.

As shown, 8-bit clusters (global and per layer) and 7-bit clusters (global only) achieve high accuracy. Regarding 7, 6, and 5-bit clusters per layer, mAP drops moderately. However, 6 and 5-bit clusters applied globally lead to dramatic accuracy loss.

Overall, the best setups correspond to 8-bit clusters, or to 7, 6 or 5-bit per layer clusters depending on the degree of accuracy that can be sacrificed for the sake of reducing energy consumption and memory bandwidth requirements.

\subsection{Bandwidth Evaluation}

\begin{table}[!h]
	\caption{Bandwidth, FPS, and energy comparison.}
	\label{tb:bwe}
	\small
	\setlength{\tabcolsep}{1.0mm} 
	\centering
	\begin{tabular}{|l||c|c|c|c|}
		\hline
		\textbf{Configuration} & \textbf{Bandwidth} & \textbf{FPS} & \multicolumn{2}{c|}{\textbf{Relative energy}} \\
		\cline{4-5}
		\textbf{             } & \textbf{GB/s     } &  & \textbf{Memory only} & \textbf{Overall} \\
		\hline
		\textbf{Baseline}         & 200.0    &  25 & 100.0\% & 100.0\% \\
		\hline
		\textbf{Clustered 8 bits} &  77.1 &  65  &  38.9\% &  48.4\% \\
		\hline
		\textbf{Clustered 7 bits} &  73.0 &  68  & 38.9\% &  48.4\% \\
		\hline
		\textbf{Clustered 6 bits} &  68.9 &  73 & 34.8\% &  45.9\% \\
		\hline
		\textbf{Clustered 5 bits} &  63.4 &  79  & 32.0\% &  42.6\% \\
		\hline
	\end{tabular}
\end{table}

Memory bandwidth requirements for the different clustering configurations as well as for the baseline YOLOv3 are provided in Table~\ref{tb:bwe}. Note that in all cases we assume that object detection must be performed at a rate of 40ms per frame (25 FPS). Under these assumptions, the baseline YOLOv3 requires 200 GB/s (199.97 to be more precise) of DRAM memory bandwidth. The particular DRAM modelled has a theoretical maximum bandwidth of 204.8 GB/s (8 channels * 3.2 GHz * 8 bytes per transfer). Hence, it could sustain the bandwidth required very tightly. Clustering, instead, allows reducing bandwidth requirements down to 63-77 GB/s (i.e., a reduction of the bandwidth required between 61\% and 68\%), therefore enabling the use of less aggressive (and cheaper) DRAM memory systems. 
Hence, we could achieve 25 FPS with a memory system delivering much lower bandwidth (e.g., a memory offering 63.4 GB/s would allow reaching 25 FPS if 5-bit clusters are used).
In other words, if we keep the same DRAM memory reaching 204 GB/s, weight clustering would allow reaching 65-79 FPS.

\subsection{Energy Evaluation}

First, we have evaluated the energy consumption of the centroids table using SRAM memories across the different configurations, hence varying the SRAM size, the number of ports, as well as the number of SRAMs (tables) needed in each case. Energy consumption of the centroids tables is below 0.1\% of the total energy in all cases. Therefore, while it is included in the estimates, we do not further discuss its contribution in the rest of this section.

Table~\ref{tb:bwe} (column corresponding to relative memory energy) provides the results for DRAM and SRAM memories consolidated. In particular, it reports their relative energy consumption with respect to the baseline configuration. As shown, weight clustering achieves large memory energy reductions in the range 61-68\%. In our evaluation, we assume that all data are word aligned (32-bit alignment) so that no weight value spans across two different 32-bit words. Therefore, 8-bit and 7-bit clustering allow storing 4 weights per word, whereas 6-bit clustering and 5-bit clustering can encode 5 and 6 weights per 32-bit word respectively. Hence, 8-bit and 7-bit clustering achieve analogous energy savings, as shown in the table.
Since the centroids tables produce negligible energy consumption, the difference between applying clustering per-layer or globally is negligible (largely below 0.1\%), so results with one decimal digit match across both configurations in all cases, and therefore, we do not report them twice.

Results also accounting for the energy consumed by FP operations are shown in the last column. Overall, clustering achieves total energy reductions in the range 52-57\%, and hence, energy consumption is between 43\% and 48\% that of the original YOLOv3 without weight clustering. Note also that the reduction between a given number of bits for clustering between All Layers and Per Layer approaches is virtually the same as discussed before.

\subsection{Tradeoff Analysis}

\begin{figure}[h!]
	\centering
	\includegraphics[width=0.75\columnwidth]{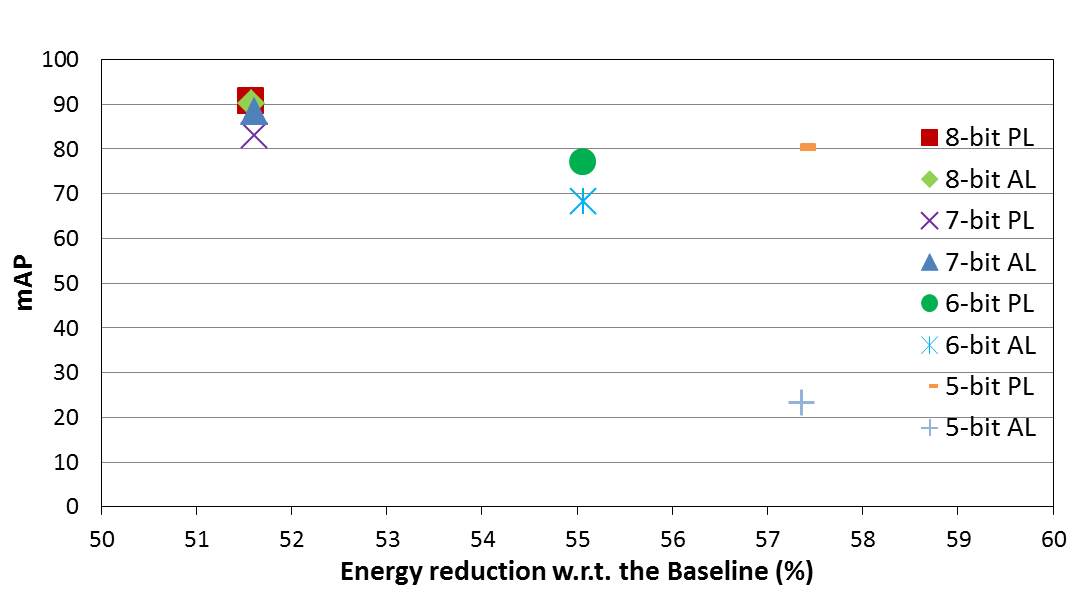}
	\caption{mAP and Energy reduction tradeoff for the different clustering configurations (Video dataset). The baseline configuration corresponds to (0\% energy reduction, 100 mAP).}
	\label{fig:map-energy-videos}
\end{figure}

\begin{figure}[h!]
	\centering
	\includegraphics[width=0.75\columnwidth]{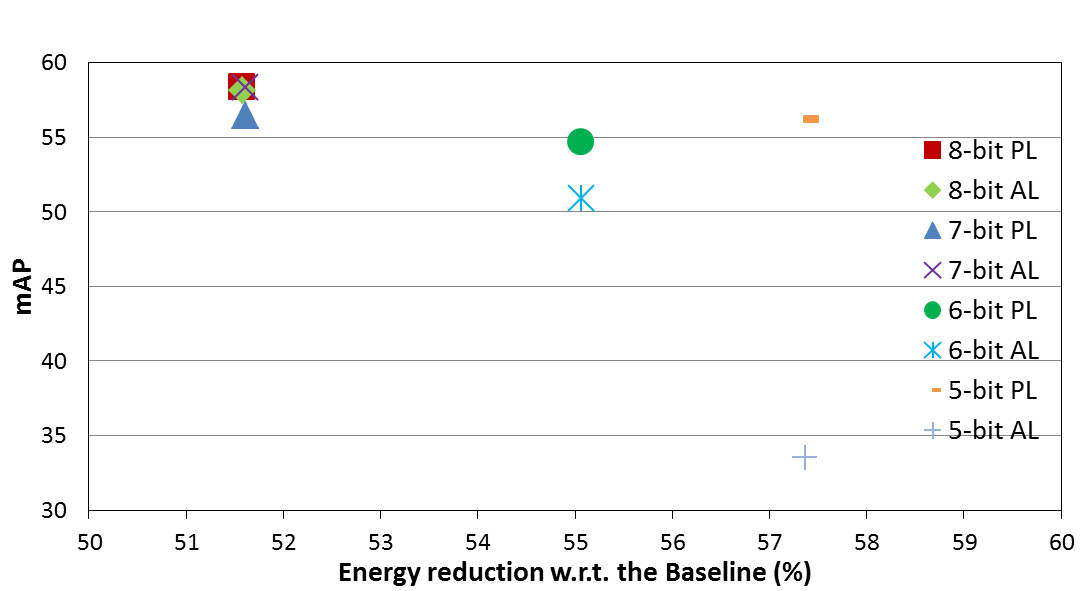}
	\caption{mAP and Energy reduction tradeoff for the different clustering configurations (Labelled dataset). The baseline configuration corresponds to (0\% energy reduction, 61 mAP).}
	\label{fig:map-energy-labelled}
\end{figure}

Figures~\ref{fig:map-energy-videos} and ~\ref{fig:map-energy-labelled} put together accuracy and energy savings in a single plot for the video dataset and labelled dataset, respectively.
In the figures, the AL abbreviation indicates clustering of \textit{All Layers}, and the PL abbreviation indicates clustering \textit{Per Layer}.
Note that, since bandwidth relates highly linearly with energy consumption, we use only energy consumption in the plots.

Looking at both plots, we can reach the following two conclusions: (1) if accuracy has more importance than energy, then 8-bit clusters, either for all layers or per layer, provide high energy savings with limited impact in accuracy. (2) However, if energy consumption is the most relevant feature to optimize, then 5-bit clusters per layer are the best choice since they deliver an additional 6\% energy reduction w.r.t. 8-bit clusters, and its accuracy is larger than that of any other configuration providing comparable energy savings.

Note that 7-bit AL provides virtually the same accuracy as 8-bit clustering for the labelled dataset. However, its accuracy is lower for the video dataset. Therefore, 8-bit clustering is a better option overall. Analyzing the results, we reach the conclusion that some configurations could be safely discarded: e.g., 7-bit PL and 7-bit AL could be discarded, since 8-bit PL and 8-bit AL provide the same energy reduction but provide higher accuracy. 6-bit PL and 6-bit AL could also be discarded, since 5-bit PL provides both, higher energy reductions and higher accuracy. Lastly, 5-bit AL can be discarded since its accuracy is not acceptable, and 5-bit PL provides the same energy reduction with an acceptable accuracy.

%% file: 5.0.Related.tex
\section{Related Work}
\label{sec:related}

Related works in the area of clustering for neural networks are abundant. The work from Han et al.~\cite{han2016deep} analyses the impact of clustering and network pruning during training and also Huffman coding after training for LeNet-300, LeNet-5, AlexNet, and VGG-16, but does not analyze the energy impact of weight clustering alone. Han et al.~\cite{fp-energy} also propose the first accelerator using clustering, pruning and Huffman coding building on top of the previous work. The work analyzes the three techniques combined and 16-bit fixed-point data to be able to reduce the model to fit it into an on-chip SRAM, consuming orders of magnitude less energy than a DRAM. This is not possible in our case since the YOLOv3 model cannot be reduced as drastically applying only clustering without sacrificing accuracy significantly. 

Choi et al.~\cite{DBLP:journals/corr/ChoiEL16} propose a similar workflow to Han et al.~\cite{han2016deep}, implementing pruning, clustering, and Huffman coding on LeNet, ResNet, and AlexNet. The work proposes the use of the Hessian-weighted k-means to perform weight quantization, showing that the Hessian-weighted k-means is a viable alternative to the traditional k-means algorithm and allows applying weight clustering to the entire model, rather than per layer as used by Han et al.~\cite{han2016deep}, with negligible accuracy loss. Wang et al.~\cite{9122495} propose an accelerator for CPU+FPGA-based heterogeneous platforms for YOLOv2 by exploiting pruning, weight-clustering, and quantization. However, this work uses the PASCAL VOC dataset to assess the accuracy of the different configurations, while our work uses a more recent version of YOLO, and provides a more complex case study of real driving videos. Moreover, our work focuses on the weight-clustering technique, providing an in-depth analysis of its hardware implications, and the accuracy and energy impact for different clustering configurations. Ye et al.~\cite{ye2018unified} is a more recent work that studies pruning and clustering in LeNet-5, AlexNet, and VGG-16 using the MNIST and Imagenet datasets. The work proposes the use of ADMM-based pruning and clustering to exploit the maximum degree of data redundancy. Their approach quantizes a portion of the weights in each training iteration, retraining the non-quantized weights, outperforming prior works. Tung et al.~\cite{8578919} propose to perform pruning and quantization in parallel during the training process, and analyze the accuracy impact in AlexNet, GoogleNet, and Resnet. Authors claim that their methodology allows recovery from premature pruning. However, their work uses uniform quantization, which consists of dividing the numerical range of weights into uniform parts, as opposed to k-means, which is a non-uniform quantization method.

Other works~\cite{gong2014compressing,app9122559} also explore K-means clustering in terms of accuracy, but do not report energy analyses. It is also worth noting that most works analyze clustering in much less complex neural networks than YOLO, and focus on the image classification domain. Because of this significant complexity difference, works such as~\cite{gong2014compressing} achieve as low as 1-bit weights, but this is not feasible with YOLOv3 when applying post-training clustering. Works such as~\cite{DBLP:journals/corr/CourbariauxB16,DBLP:journals/corr/CourbariauxBD15} also propose binary weights during training. Other works consider other forms of quantization such as integer quantization~\cite{wu2020integer} or uniform quantization~\cite{DBLP:journals/corr/ZhouYGXC17,DBLP:journals/corr/abs-1806-08342}.

In our case study, we assess the impact of post-training weight clustering in terms of accuracy and energy for a large case study such as YOLOv3. The main contribution of our work is the in-depth accuracy and energy analysis of weight clustering for a complex camera-based object detection system, covering the gap of previous works. To the best of our knowledge, our work is the first one to provide such a complex case study, thus helping designers make the right choice for the design of their accelerators and providing hints on how to evaluate accuracy, energy and bandwidth at scale.

%% file: 6.0.Conclusions.tex
\section{Conclusions and Future Work}
\label{sec:concl}

Weight clustering is a promising solution to decrease the memory bandwidth requirements and energy consumption of DNNs. However, it has been evaluated in limited examples so far, hence providing insufficient information of its real benefits and tradeoffs.
This paper presents an at-scale case study considering different flavors of weight clustering, the main characteristics of an appropriate accelerator with the right dataflow, a relevant application such as YOLOv3 for camera-based object detection, and both, reference datasets as well as real driving condition videos. 
Our results provide relevant information in terms of memory bandwidth and energy consumption, showing that energy gains range between 61\% and 68\% for memory, and between 52\% and 57\% for the whole framework with different degrees of accuracy.
Our results also allow concluding that, if applied aggressively (e.g., 5-bit clusters), weight clustering must be better applied in a per-layer basis to mitigate accuracy degradation.

As part of our ongoing work, we are contributing to the design of a systolic array accelerator that adopts the findings in our work with the aim of offering it as open source (with permissive licenses). Such accelerator will be further integrated in RISC-V open source SoCs.